\documentclass{iau}

\usepackage{amsmath}
\usepackage{graphicx}
\usepackage{multirow}
\usepackage{natbib}

\newcommand\apj{{\it{ApJ}}}
\newcommand\solphys{{\it{SoPh}}}
\newcommand\jgr{{\it{JGR}}}

\begin{document}

\lefttitle{Joshi {\it al.}}
\righttitle{CME and DH Type II radio bursts }

\jnlPage{1}{7}
\jnlDoiYr{2021}
\doival{10.1017/xxxxx}
\volno{388}
\pubYr{2024}
\journaltitle{Solar and Stellar Coronal Mass Ejections}

\aopheadtitle{Proceedings IAU Symposium}
\editors{N. Gopalswamy,  O. Malandraki, A. Vidotto \&  W. Manchester, eds.}

\title{DH type II radio bursts during solar cycles 23-25: Origin and association with solar eruptive events}

\author{Bhuwan Joshi$^1$, Binal D. Patel$^{1}$, Kyung-Suk Cho$^{2}$, and Rok-Soon Kim$^{2}$}
\affiliation{{}$^1$Udaipur Solar Observatory, Physical Research Laboratory, Udaipur 313001, India\\email: \email{bhuwan@prl.res.in}}
\affiliation{{}$^2$Space Science Division, Korea Astronomy and Space Science Institute, Daejeon 34055, Republic
of Korea}

\begin{abstract}
We analyses occurrence of DH type II solar radio bursts spanning over solar cycles 23-25 during which a total of 590 DH type II bursts are reported with confirmed 568 and 462 cases of associated CME and flares, respectively. We find short-term yet important differences in DH type II activity when the data is examined in terms of event counts and their durations, e.g., temporal shift in the peak activity during cycle 24 and variation in the growth rate of the activity level during cycle 25. For an in-depth exploration, DH type II bursts are classified in 3 categories based on their end-frequencies: Low-, Medium-, and High- Frequency Groups (LFG, MFG, and HFG, respectively). The HFG category is the most populous ($\approx$47\%) while the LFG category occupy about a quarter of the events ($\approx$24\%). The LFG events show a clear inclination toward fastest CMEs and X-class flares with a quarter of events exhibiting end frequency below 50~MHz.

\end{abstract}


\maketitle

\section{Introduction}

It is well recognized that the powerful CMEs generate shocks in the coronal and interplanetary medium. In radio dynamic spectrum, these shocks are identified as type II bursts. Depending on the energetics of CMEs, the type II bursts are observed in metric (m;30 MHz~$\leq$~f~$\leq$~300 MHz), decametric–hectometric (DH; 300 kHz~$\leq$~f~$\leq$~30 MHz), and
kilometric (km; 30 kHz~$\leq$~f~$\leq$~300 kHz) wavelength domains. The extension or origin of type II radio bursts in the DH domain implies the cases of stronger MHD shocks propagating from the inner corona and entering the interplanetary (IP) medium \citep{Gopal2001,Gopal2005JGRA}. Hence the study of
shocks in DH domain, together with their associated CME-flare events, becomes extremely
important to infer not only the propagation characteristics of CMEs but also to develop their
forecasting tools \citep{Reiner2007,Joshi2018,Syed2019}. Contextually, the CME associated with type II radio bursts in metric and/or DH wavelength domain are termed as radio loud CMEs \citep{Michalek2007,Gopal2008}.

\begin{table}[h!]
\centering
\caption{Occurrence of DH type II radio bursts from January 1996 to June 2023, covering solar cycles 23-25, along with the identified solar events, i.e., CMEs and flares, occurred in association with the reported bursts.}\label{Tab:counts-total}
 {\tablefont\begin{tabular}{@{\extracolsep{\fill}}clll}
    \midrule
    Solar Cycle &No. of Events\\
    \cline{2-4}
    &{DH type II}&\multicolumn{2}{c}{Identified solar events} \\
    \cline{3-4}
     &(\% out of total)& CME & Flare\\
    \midrule
     23 & 335 (57\%) & 320 & 275\\
     24 & 180 (30\%) & 173 & 127\\
     25 & 75 (13\%)& 75 & 60\\
     \midrule
     23+24+25    & 590 & 568 & 462\\
     \midrule
     \end{tabular}}
\end{table}

In this paper, we present a statistical study of DH type II bursts occurred during solar
cycles 23, 24, and 25. For the present analysis, we have obtained data from the following sources: (1) Wind/WAVES Type-II Burst Catalogue (\url{https://cdaw.gsfc.nasa.gov/CME_list/radio/waves_type2.html}), (2) Solar and Heliospheric Observatory (SOHO) Large Angle and Spectroscopic Coronagraph (LASCO) CME Catalogue(\url{https://cdaw.gsfc.nasa.gov/CME_list/}). This work highlights the importance of analysing DH type II bursts in terms
of their end frequencies \citep{Patel2021,Patel2022}. Notably, the end frequencies of these IP bursts directly relate to the heliocentric distance up to which a shock can survive and, therefore, have implications
in exploring the energetics and propagation characteristics of CMEs. 

\section{Occurrence and frequency-dependent characteristics}

\subsection{DH type II bursts during cycles 23-25}

\begin{figure}
\center
\includegraphics[scale=.6]{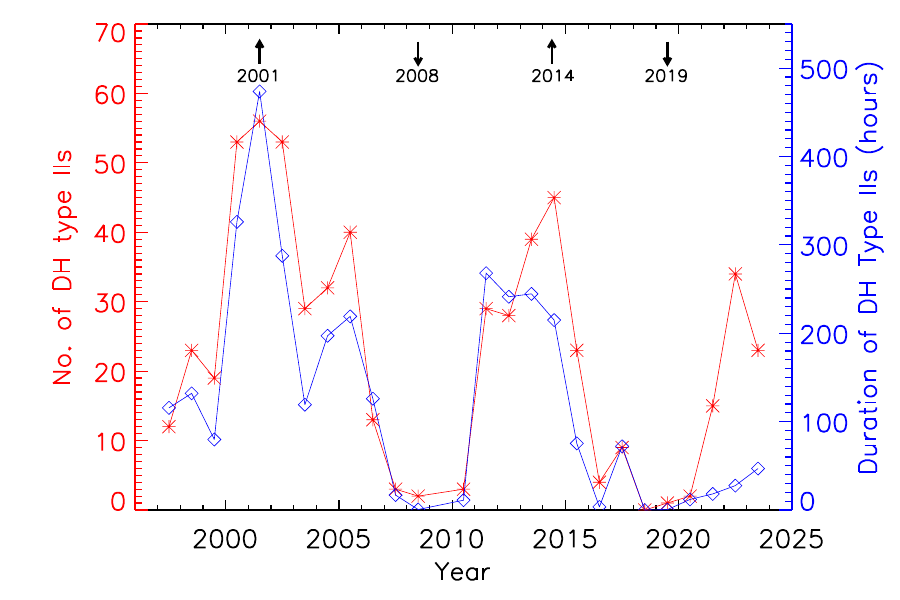}
\caption{Annual occurrence of DH type II radio bursts from January 1996 to June 2023, covering solar cycle 21, 22, and rising phase of cycle 23 in terms of counts and cumulative durations over a year. Up and down arrows represent the maximum and minimum years of solar cycles.}
\label{Fig:Yearly-plot}
\end{figure}

\begin{figure}
\center
\includegraphics[scale=.8]{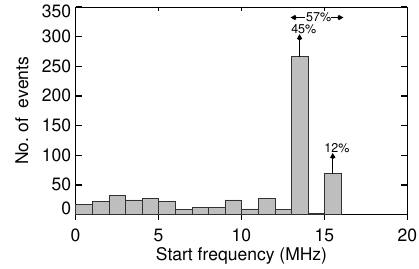}
\caption{Distribution of start frequencies of DH type II radio bursts for a total of 590 events of our data set.}
\label{Fig:type2_start_freq}
\end{figure}

A comparison of occurrence of DH type II radio bursts during solar cycles 23 (January 1996--December 2008), 24 (January 2009--Dec 2017, and 25 (January 2018--June 2023) shows that cycle 23 produced about twice number of events than cycle 24 (Table~\ref{Tab:counts-total}). The solar cycle 25 is still running, however, in the beginning phase of $\approx$5 years, considered here, it has exhibited a deficit in DH type II activity compared to the previous two cycles along with a slow rise. To understand the activity level of DH type II radio bursts with the evolution of solar cycles, we characterize the type II activity in terms of their total yearly counts and cumulative durations (Figure~\ref{Fig:Yearly-plot}). Both indicators reflect an obvious cyclic behaviour of $\approx$11 years with some noteworthy differences: the maximum activity period for the two parameters differs for cycle 24, and type II event counts show a steep growth during 2021-2023 while the cumulative durations indicates a slow and gradual rise. To have a clear understanding of the observed DH type II events in terms of the frequency at which the bursts originated or first detected, we present a histogram of the starting frequencies in Figure~\ref{Fig:type2_start_freq}. Here we clearly notice that the histogram bars representing the limiting frequency range of Wind/WAVES (13–14 MHz) and SWAVES (15–16 MHz) populate 45\% and 12\% events,
respectively. From this statistics, it is clear that a sizeable population of DH type II events exist as an extension of type II emission starting at metric frequencies. 

\subsection{Frequency-dependent categories and their characteristics}

  \begin{figure}
    \center
    \includegraphics[scale=.45]{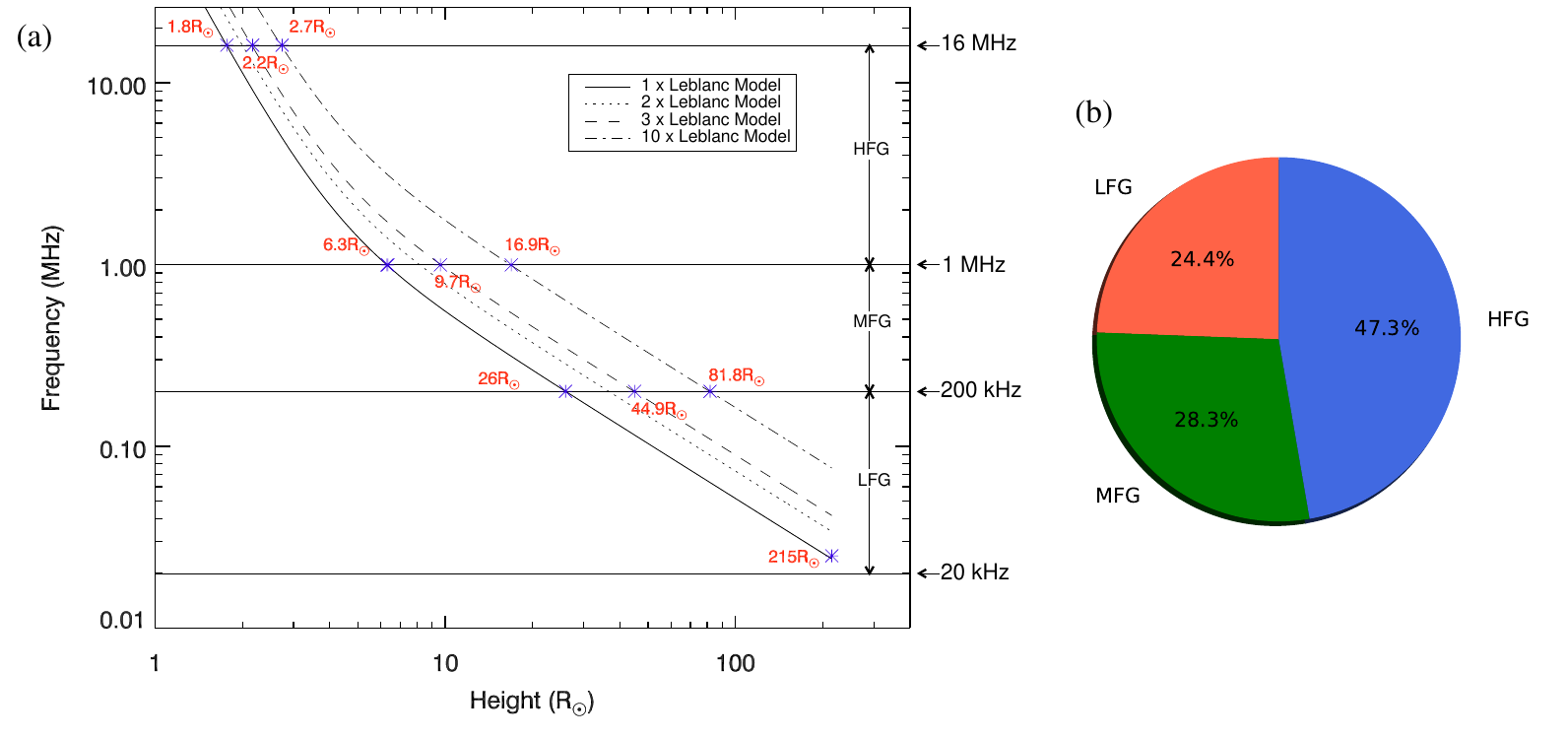}
    \caption{(a) Variation of heliocentric burst height (R$_\odot$) with plasma frequency estimated from Leblanc electron density model. (b) Pie chart showing the fraction of LFG (red),
MFG (blue), and HFG (green) DH type II radio bursts occurred during cycles 23-25.}
    \label{Fig:Leblanc_frequecy_height}
  \end{figure}
  
The characteristics feature of type II burst is slow frequency drift toward lower frequencies which is physically attributed to the burst driver moving outward from the solar atmosphere to larger heliocentric distances with ever decreasing electron densities. This frequency–height relationship of the type II burst is illustrated using the atmospheric density model of \cite{Leblanc1998} in Figure~\ref{Fig:Leblanc_frequecy_height}. Contextually, we note that the height estimations from the atmospheric density model provide a “coarse” estimation only as the density of the medium, through which shock is propagating, evolves with the spatially and temporally varying coronal and interplanetary conditions. Thus, rationally, we consider a multiplier (or suitable “fold”) to the basic atmospheric density model. 

\begin{table}[h!]
\centering
	\caption{Occurrence of DH type II events during Solar Cycles 23, 24, and 25 for LHG, MFG, and HFG categories.}
	{\tablefont\begin{tabular}{@{\extracolsep{\fill}}ccccc}

	\hline
		Solar Cycle & 	\multicolumn{4}{c} {Number of events} \\
      \cline{2-5}

          & LFG  & MFG & HFG & Total\\
        &\textit{f} $ \leq$ 200 kHz & 200 kHz $<$ \textit{f} $\leq$ 1 MHz & 1 MHz $<$ \textit{f} $\leq$16 MHz&\\
          \hline
           23 & 81(24\%) & 96 (29\%)&158 (47\%) &335 (57\%)\\
           24 & 57 (32\%) & 55 (30\%) & 68 (38\%) &180 (31\%)\\
           25 & 6 (8\%)   &  16 (21\%) & 53 (71\%) & 75 (13\%) \\
           23+24+25 &  144 (24.4\%)& 167 (28.3\%) & 279 (47.3\%)&590\\
          \hline

\end{tabular}}
\label{Tab:counts_LFG_MHG_HFG}
\end{table}

In Figure~\ref{Fig:Leblanc_frequecy_height}a, we draw frequency-height relation for 1-, 2-, 3-, and 10-fold Leblanc models to incorporate density variations of upto an order. The frequency–height relationship suggests that the observing window of 16 MHz–20 kHz essentially represents a large heliocentric distance from $\approx$2R$_\odot$ to 1 AU. Therefore, to explore the characteristic of DH type II radio bursts and associated solar eruptive events, we divide this vast frequency range into three groups: Low-Frequency Group (LFG; 20 kHz~$\leq$~f~$\leq$~200 kHz), Medium-Frequency Group (MFG; 200 kHz~$\leq$~f~$\leq$~1 MHz), and High Frequency Group (HFG; 1 MHz~$\leq$~f~$\leq$~16 MH. Figure~\ref{Fig:Leblanc_frequecy_height}a readily shows that our end-frequency classification essentially provides a quantitative estimation on the distance up to which a shock can survive. We further note that HFG and MFG events represent shocks terminating within the lower and upper coronal heights (9.7R$_\odot$~and~44.9R$_\odot$, as per three-fold Leblanc model), respectively. On the other hand, the events under LFG group represent shock travelling in the interplanetary medium (beyond $\approx$45R$_\odot$). 

\begin{figure}
    \center
    \includegraphics[scale=.65]{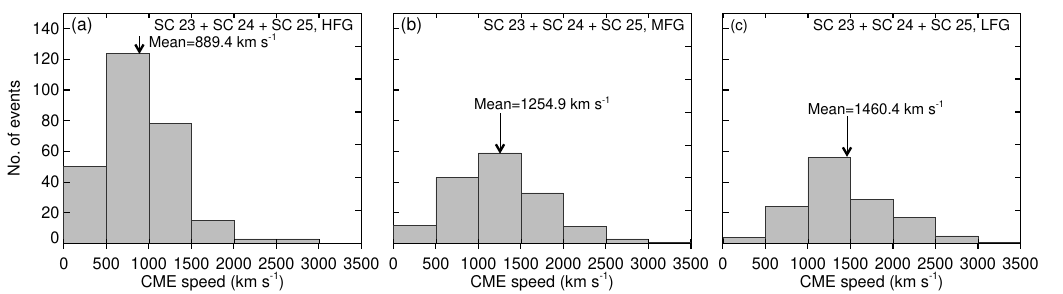}
    \caption{Histogram showing the distribution of linear speed of CMEs (within LASCO field-of-view) for events of HFG (panel a), MFG (panel b), and LFG (panel c) categories.}
    \label{Fig:hist_CME_speed}
  \end{figure}

In Table~\ref{Tab:counts_LFG_MHG_HFG}, we list the counts for LFG, MFG, and HFG events for solar cycle 23, 24, and 25. The same has been pictorially shown in a pie chart in Figure~\ref{Fig:Leblanc_frequecy_height}b. In Figure~{\ref{Fig:end_freq}a, we present the histogram of ending frequency for DH type II events which reveal 49\% of events to lie below 1 MHz frequency bin. We further show the distribution of ending frequencies from 1 MHz down to 20 KHz in Figure~\ref{Fig:end_freq}b to explore the occurrence of LFG and MFG events exclusively. In Table~\ref{Tab:lt_50MHz_events}, we specifically give number of type II bursts that end beyond 50~MHz as this frequency roughly corresponds to heliocentric distance of 0.5 AU (see Figure~\ref{Fig:Leblanc_frequecy_height}a); We find cycle 24 to be highly deficient in comparison to cycle 23 (12\% versus 37\%) while rise phase of cycle 25 completes lacks such events. 
     
 \begin{figure}
    \center
    \includegraphics[scale=.75]{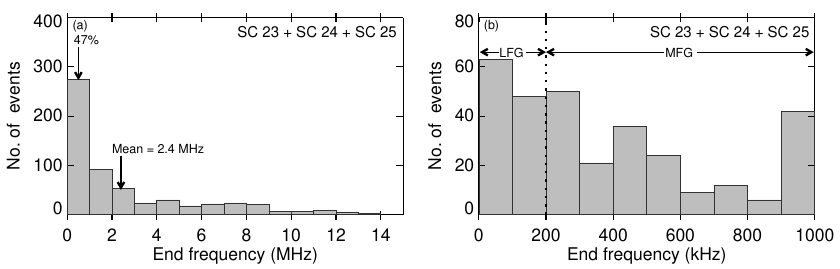}
    \caption{(a) Distribution of DH type II radio bursts in terms of their end frequencies. (b) Distribution of end frequencies exclusively for LFG and MFG events.}
    \label{Fig:end_freq}
  \end{figure}  
  
\begin{table}[h!]
\centering
\caption{Occurrence of DH type II events having end frequency $<$50 kHz.}\label{Tab:lt_50MHz_events}
 {\tablefont\begin{tabular}{@{\extracolsep{\fill}}ccc}
    \midrule
    Solar Cycle &No. of Events & \%\\
                & (total events)& \\
    
    \cline{2-3}
    23 &  30 (81) & 37\% \\
    24 &  7 (57) & 12\%\\
    25 & 0 (6) & 0\\
    23+24+25 & 37 (144) & 26\\
   \midrule
     \end{tabular}}
\end{table}

\section{Association with CMEs and flares}

In Figure~\ref{Fig:hist_CME_speed}, we present the distribution of linear speed of CMEs within LASCO field-of-view associated with DH type II radio bursts for HFG, MFG, and LFG categories. The histograms clearly reveal a larger difference of $\approx$570~km~s$^{-1}$~between the mean CME speeds for two limiting cases i.e., LFG and HFG categories. Further,
as expected, the mean speed for MFG events lies between the LFG and HFG classes. To investigate whether the CME speeds for HFG, MFG, and LFG categories belong to the same distribution or not (i.e., if the difference in mean
speeds of CMEs for the three successive categories is statistically significant), we perform the two-sample Kolmogorov–Smirnov test \citep[(K-S test;][]{Press1992} and the results of the test is presented in Table~\ref{Tab:ks_test}}. The two-sample K-S test suggests that the difference in the mean speed of CMEs between the events of three categories of DH type II radio bursts is statistically highly significant.

\begin{figure}
    \center
    \includegraphics[scale=.65]{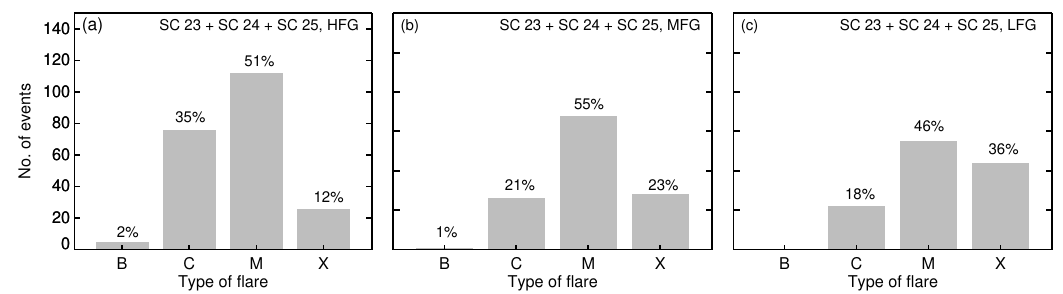}
    \caption{Histograms showing the distribution of flare counts of GOES classes (B, C, M, and X) associated with
DH-type II radio bursts of HFG (panel a), MFG (panel b), and LFG (panel c) categories.}
    \label{Fig:hist_flare_class}
  \end{figure}

In Figure~\ref{Fig:hist_flare_class}, we present the histogram showing the association of flares of different GOES classes (viz B, C, M, and X) with DH type II radio bursts. We find that the highest number of
events are associated with the M class flares ($\approx$55\%) for all the categories. We further find that the fraction of occurrence of X-class flares becomes almost double as we go from HFG to LFG categories (12, 23, and 36\%, respectively).

\begin{table}[h!]
\centering
\caption{The results of the two-sample K-S test assessing the observed difference between mean CME speeds of HFG, MFG, and LFG categories. N$_{1}$ and N$_{2}$ represent the number of events for respective categories. The K-S test statistics and probability are denoted by D and P.}\label{Tab:ks_test}
 {\tablefont\begin{tabular}{@{\extracolsep{\fill}}cccccc}
    \midrule
    Category&$\Delta$V &\multicolumn{2}{c}{Sample Size} & D  & P\\
    \cline{3-4}
    & (km~s$^{-1}$)&N$_1$ & N$_2$& &\\
     \midrule
    $\Delta$V$_{\rm {MFG-HFG}}$ & 365 & 162 & 273& 0.33 & 2.5$\times$10$^{-10}$\\
    $\Delta$V$_{\rm {LFG-MFG}}$ & 206 & 136 & 162& 0.205 & 0.0032\\
    $\Delta$V$_{\rm {LFG-HFG}}$ & 571 & 136 & 273 & 0.515 & 5.7$\times$10$^{-22}$\\
     \midrule
     \end{tabular}}
\end{table}

\section{Conclusions}
The present study make a detailed survey of DH type II radio bursts occurred during solar cycles 23-25 involving a total of 590 DH type II bursts from January 1996 to June 2023, with confirmed 568 and 462 cases of associated CME and flares, respectively. In the following, we summarize the main results of the paper:

\begin{enumerate}

\item The occurrence rate of DH type bursts follow a periodic behaviour with the phases of solar cycle. However, subtle but important differences are seen when the data is examined in terms of event counts and event durations. This result has important implication in understanding the origin of solar activity and its coronal and heliospheric consequence.

\item Our approach is to classify radio loud CMEs based on the end-frequencies of the corresponding DH type II radio bursts. Accordingly, three categories of CMEs have been considered viz. Low-Frequency Group (LFG; 20 kHz~$\leq$~f~$\leq$~200 kHz), Medium-Frequency Group (MFG; 200 kHz~$\leq$~f~$\leq$~1 MHz), and High Frequency Group (HFG; 1 MHz~$\leq$~f~$\leq$~16 MH. This end-frequency classification plausibly describes the heliocentric distance up to which a shock can survive. 

\item The HFG category is the most populous, occupying almost half of the events of the entire sample ($\approx$47\%). The LFG category (f~$\leq$~200 kHz), most important for space weather perspective, occupy a quarter of the events ($\approx$24\%).

\item Within the LFG class, $\approx$25\% events exhibiting end frequency below 50~MHz, are of special interest as this frequency corresponds to a heliocentric distance of $\approx$0.5~AU. Notably, cycle 24 drastically lacks such events in comparison to cycle 23 (12\% versus 37\%). In cycle 25, we could not find any events of this category within the observing period (January 2018--June2023).

\item The LFG events show a clear inclination toward fastest CMEs and largest X-class flares.
\end{enumerate}

In summary, the present study provides us insights about the statistical properties of DH type II radio bursts during cycles 23-25 but also widens our understanding about the conditions related to their origin in the near-Sun region and survival in the corona and interplanetary medium. Our results indicate that the three-element classification of radio loud CMEs -- HFG, MFG, and LFG -- has important implications. In particular, the HFG events imply propagation of the shock inside the complex magnetic environment and within faster solar wind condition of the low corona, whereas, the events of LFG category, representing the shock travelling from the corona to the interplanetary medium, are rather intriguing for understanding the space weather phenomena.

The research work at the Physical Research Laboratory (PRL) is funded by the Department of Space, Government of India. We gratefully acknowledge the WIND/WAVES type II burst catalog and LASCO CME catalog. We further acknowledge the SOHO, STEREO, GOES, and Wind missions for their open data policy. BJ thanks the organizers for providing the local hospitality.


\end{document}